\documentstyle[12pt]{article}

\newcommand{\slp}{\raise.15ex\hbox{$/$}\kern-.57em\hbox{$\partial$}}
\newcommand{\sla}{\raise.15ex\hbox{$/$}\kern-.57em\hbox{$a$}}
\newcommand{\slA}{\raise.15ex\hbox{$/$}\kern-.57em\hbox{$A$}}
\newcommand{\slB}{\raise.15ex\hbox{$/$}\kern-.57em\hbox{$B$}}
\newcommand{\slb}{\raise.15ex\hbox{$/$}\kern-.57em\hbox{$b$}}
\newcommand{\slW}{\raise.15ex\hbox{$/$}\kern-.57em\hbox{$W$}}

\newcommand{\be}{\begin{equation}}
\newcommand{\ee}{\end{equation}}
\newcommand{\bear}{\begin{eqnarray}}
\newcommand{\ear}{\end{eqnarray}}

\newcommand{\Li}{\cal L}
\newcommand{\D}{\cal D}

\begin{document}
\begin{flushright}
HD--THEP--95--31\\
\end{flushright}
\quad\\
\vskip1.5cm

\begin{center}

\vspace{.5cm}
{\LARGE BRST  Analysis of QCD$_2$ as a Perturbed}

\vspace{.3cm}
{\LARGE WZW Theory}

\vspace{1cm}
D. C. Cabra\footnote{CONICET, Argentina}\\
{\normalsize\it
Departamento de F\'\i sica, Universidad Nacional de La Plata,}\\
{\normalsize\it
C.C.~67, (1900) La Plata, Argentina.}\\
\vspace{1cm}
%
K. D. Rothe\\
{\normalsize\it
Institut  f\"ur Theoretische Physik
Universit\"at Heidelberg}\\
{\normalsize\it Philosophenweg 16, D-69120 Heidelberg,
Germany}\\
%
%
\vspace{1cm}
F.A. Schaposnik\footnote{Investigador CICBA}\\
{\normalsize\it
Departamento de F\'\i sica,
         Universidad Nacional de La Plata}\\
 {\normalsize\it C.C. 67, (1900) La Plata, Argentina}
\date{}
\end{center}
\pagestyle{empty}
\vspace{1.5cm}
\begin{abstract}\noindent
Integrability of Quantum Chromodynamics in 1+1 dimensions has
recently been suggested by formulating it as a perturbed
conformal Wess-Zumino-Witten theory.
The present paper further elucidates this formulation,
by presenting a
detailed BRST analysis.
\end{abstract}
\newpage
\pagestyle{plain}
\section{Introduction}
As the non-Abelian extension of the exactly soluble Schwinger model
\cite{Schwi}, Quantum Chromodynamics in $1+1$ dimensions ($QCD_2$) has
received much attention in the past 25 years. Its equivalent bosonic
description was, however, fully understood only

as late as 1984, with the
work of Polyakov-Wiegmann \cite{PolW} and Witten \cite{Wit}.

The basic idea for arriving at an equivalent bosonic description
of QCD$_2$ is to perform a change of variable which decouples the
fermions from the gauge field (decoupled picture). This idea,
explored by \cite{RS} in the bosonization of the Schwinger model
\cite{Schwi}, has been successfully applied in a number of papers on
QCD$_2$ \cite{GSS}, \cite{FNS}.
For a general review the reader is referred to
\cite{AAR}. Following this procedure one arrives at an effective
action involving a conformally
invariant Wess-Zumino-Witten [WZW] functional
plus the Yang-Mills action.

By regarding the Yang-Mills action of QCD$_2$ as a perturbation
of a conformally invariant WZW theory of positive and negative

level WZW fields, as well as ghosts, in the decoupled picture,
Abdalla and Abdalla \cite{AA}
obtained an infinite set of conservation laws.
In their description a number
of first and second class constraints emerged the significance and
role of which, however, remained unclear.

It is the objective of this note
to clarify this and related aspects of their approach, by performing a
detailed BRST analysis of $QCD_2$ following the path-integral
formulation described above.

\section{Local, decoupled formulation of $QCD_2$}
\setcounter{equation}{0}
The  partition function of $QCD_2$ is given by
\be\label{1.1}
Z=\int[{\D}A_\mu]\int{\D}\psi{\D}\bar\psi \exp[{i(S_{YM}+S_F)}]
\ee
where $S_{YM}$ and $S_F$ are the Yang-Mills and fermionic
action, respectively\footnote{Our conventions are $\partial_\pm=
\partial_0\pm\partial_1,A_\pm=A_0\pm A_1$,
$F_{\mu\nu}=t^aF^a_{\mu\nu}$, etc., with the normalization
$tr t^at^b=\delta^{ab}$ for the
hermitian generators in the fundamental representation,

and the commutation relation $[t^a,t^b]=ifabc\ t^c$,

with $fabcfabd=\frac{C_V}{2}\delta cd$.
We follow in general the notation and conventions of ref. \cite{AA}.}
\be\label{1.2}
S_{YM}=-\frac{1}{4}\int d^2xtr(F_{\mu\nu}F^{\mu\nu})\ee
\bear\label{1.3}
S_F&=&\int d^2x\bar\psi(i\slp+e\slA)\psi\\
&=&\int d^2x\left\lbrace \psi_1^\dagger(i\partial_++eA_+)\psi_1+\psi_2^\dagger
(i\partial_-+eA_-)\psi_2\right\rbrace\nonumber\ear
and $[{\D} A_\mu]$ stands for the measure including gauge fixing.
It will be convenient to work in the light cone gauge $A_+=(A_0+A_1)=0$. We
implement this gauge
in terms of a Lautrup-Nakanishi-Lagrange multiplier field $B$,
\be\label{1.4}
Z=\int{\D}A_\mu\int{\D}\psi{\D}\bar\psi\int{\D}B\int{\D}
b_-{\D}c_- \exp({iS_{GF}})
\ee
with
\be
\label{1.5}
S_{GF}=S_{YM}+S_F+S_{gh}+\int d^2x tr(BA_+)
\ee
where  $S_{gh}$ is the ghost action
\be\label{1.6}
S_{gh}=\int d^2x tr(b_-i{\D}_+c_-)\ee
with
\be\label{1.7}
{\D}_\pm=\partial_\pm-ie[A_\pm,\quad]\ee
the covariant derivative in the adjoint representation.

The gauge-fixed action (\ref{1.5}) is invariant under the BRST
tranformation
\bear\label{1.8a}
\delta A_\mu&=&\epsilon\frac{1}{e}{\D}_\mu c_-\nonumber\\
\delta\psi_\alpha&=&\epsilon c_-\psi_\alpha\nonumber\\
\delta c_-&=& \epsilon\frac{1}{2}\{c_-,c_-\}\nonumber\\
\delta b_-&=& \epsilon\frac{1}{e} B\nonumber\\
\delta B&=&0\ear
with $\epsilon$ a Grassman-valued infinitesimal parameter.

Performing the integration over $B$ and $A_+$, we obtain the following
partition function
\be
\label{1.9}
Z=\int{\D}A_-\int{\D}\psi_1^{(0)}D\psi_1^{\dagger(0)}\int{\D}
\psi_2{\D}\psi_2^\dagger\int{\D}b^{(0)}_-{\D}c^{(0)}_-
\exp\left({iS_{GF}}\right)
\ee
with the corresponding gauge-fixed Lagrangian
\bear\label{1.10a}
{\Li}_{GF}&=& tr\frac{1}{8}(\partial_+A_-)^2 + \psi^{(0)\dagger}_1i
\partial_+\psi^{(0)}_1\nonumber\\
&&+\psi^\dagger_2iD_-\psi_2+tr \left( b_-^{(0)}i\partial_+c^{(0)}_-\right)
.\ear
We have denoted the fields obbeying a free-field dynamics
by a superscript ``(0)'', and $D_-$ is the covariant derivative

\be\label{1.10b}
D_\pm=\partial_\pm-ieA_\pm\ee
with $A_\pm$ in the fundamental representation.

The classical action $S_{GF}$ is invariant under the BRST transformation
\bear\label{1.11}
\delta A_-&=&-i \epsilon\frac{1}{e}{\D}_- c_-^{(0)}\nonumber\\
\delta\psi_1^{(0)}&=&\epsilon c_-^{(0)}\psi_1^{(0)}\nonumber\\
\delta \psi_2&=&\epsilon c_-^{(0)}\psi_2\nonumber\\
\delta c_-^{(0)}&=& \epsilon\frac{1}{2}
\{c_-^{(0)},c_-^{(0)}\}\ear
\bear\label{1.12}
\delta b^{(0)}_-&=& \epsilon\frac{1}{e^2}{\D}_-(\partial_+A_-)
+\epsilon\psi_1^{(0)}\psi_1^{(0)\dagger}\nonumber\\
&&+\epsilon\{b_-^{(0)},c_-^{(0)}\}\ear
These transformation laws can be obtained from (\ref{1.8a}) by eliminating
$A_+$ and $B$ using
the equations of motion
\bear\label{1.8b}
A_+&=&0\nonumber\\
B^a&=&-\frac{\delta S_{YM}}{\delta A^a_+}-e\psi^\dagger_1
t^a\psi_1+ie f_{abc}b_-^bc^c_-\ear

Regarding (\ref{1.11}) as a classical transformation, it is a symmetry
of the gauge-fixed Lagrangian (\ref{1.10a}). In order to analyse the
corresponding symmetry at the partition function level, one has
however to take into account the following facts:
\vspace{ 0.3 cm}

\noindent i) The fermionic path-integral in (\ref{1.4}) (i.e., before
fixing the gauge to $A_+ = 0$) leads to
\begin{equation}
Z_{fer} = det (i\partial_+ + e A_+) det (i\partial_- + e A_-)
\times \exp ( - i\frac{e^2}{4\pi} \int d^2x A_+ A_- )
\label{a}
\end{equation}
The last term in the r.h.s. of (\ref{a}) has been added,
exploiting regularization ambiguities, so as to ensure gauge
invariance \cite{PolW}.
\vspace {0.2 cm}

\noindent ii) Analogously, the  determinant arising from integration
of ghosts with action (\ref{1.6}) has to be adjusted with the
same type of $A_+A_-$ counterterm,

\begin{equation}\label{2.15}
Z_{ghosts} =  det^{Adj} (i\partial_+ + e A_+) \times
\exp ( - i \alpha \int d^2x A_+ A_- )
\label{b}
\end{equation}
Since, as it is well-known \cite{PolW}, the determinant in the adjoint

representation is
related with that in the fundamental  through
the Casimir $C_V$ (see \cite{FNS} for details),
consistency of the regularization implies that
$\alpha$ should be chosen as $\alpha = ({e^2}/{4\pi}) C_V$.

Taking into account the i) and ii), the transformation law for
the ghost field $b_-^{(0)}$ takes the form

\be\label{1.13}
\delta b_-^{(0)}=-\frac{\epsilon}{e}{\D}_-(\partial_+A_-)+
\epsilon\psi_1^{(0)}\psi_1^{(0)\dagger}+\epsilon\{b_-^{(0)},
c_-^{(0)}\}-\epsilon e\frac{(1+C_V)}{4\pi}A_-\ee
The last term in the r.h.s. of (\ref{1.13}) arises from the
$A_+A_-$ counterterms as discussed above. The
transformations (\ref{1.11}) and (\ref{1.13}) now represent
a symmetry transformation of the partition function associated with
(\ref{1.10a}).

We now go to a new set of variables such that
the partition function factorizes in terms of decoupled
fields, by writing
\bear\label{1.14}
A_-&=&\frac{i}{e}V\partial_-V^{-1}\nonumber\\
\psi_2&=&V\psi_2^{(0)}\ear
For the corresponding transformation of the integration
measure one has

\bear\label{1.15}
{\D}A_-&=& \int{\D}b_+{\D}c_+ \exp\left({i\int d^2x\ tr\ b_+
i{\D}_-c_+}\right) \times {\D}V \nonumber\\
&=& \exp({-iC_V\Gamma[V]})\int{\D}b_+^{(0)}{\D}c_+^{(0)} \nonumber\\
& &
\exp({i\int d^2x\ tr\ b_+^{(0)}i\partial_-c_+^{(0)}})
\times {\D}V
\ear
\be\label{1.16}
{\D}\psi_2{\D}\psi^\dagger_2={\D}\psi_2^{(0)}{\D}\psi_2^{(0)\dagger}
\exp({-i\Gamma [V]})\ee
where $\Gamma[g]$ is the Wess-Zumino-Witten functional \cite{Wit}
\be\label{1.16a}
\Gamma[g]=\frac{1}{8\pi}\int d^2x\ tr\ \partial_\mu g^{-1}\partial^\mu
g+\frac{1}{12\pi}\int d^3y\epsilon^{\alpha\beta\gamma}tr[g^{-1}\partial
_\alpha g g^{-1}\partial_\beta g^{-1}\partial_\gamma g]\ee
with the remarkable property (see \cite{AAR} for details)
\bear\label{2.19a}
\delta\Gamma[g]&=&\frac{1}{4\pi}\int d^2x\ tr g\delta
g^{-1}\partial_+(g\partial_-g^{-1})\nonumber\\
&=&\frac{1}{4\pi}\int d^2x\ tr g^{-1}\delta g\partial_-
(g^{-1}\partial_+g)\ear

In terms of the new variables, the partition function
reads
\be\label{1.17}
Z=Z_F^{(0)}Z_{gh}^{(0)}Z_V\ee
where
\be\label{1.18}
Z_V=\int{\D}V \exp\left\{{-i(1+C_V)\Gamma[V]
+\frac{i}{8e^2}\int d^2x tr[\partial_+(Vi\partial_-V^{-1})]^2}\right\}
\ee
and
\be\label{1.19}
Z^{(0)}_F=\int{\D}\psi^{(0)}{\D}\bar\psi^{(0)}
\exp\left({i\int d^2x\bar\psi^{(0)}i{\raise.15ex\hbox{$/$}\kern-.57em
\hbox{$\partial$}}\psi^{(0)}}\right)
\ee
\be\label{1.20}
Z_{gh}^{(0)}=\int{\D}b_\pm^{(0)}Dc_\pm^{(0)}
\exp[{i\int d^2x \ tr\left(b_+^{(0)}i\partial_-c_+^{(0)}+b_-^{(0)}
i\partial_+c_-^{(0)}\right)}]
\ee
Notice that the WZW action enters in (\ref{1.18})
with negative level $-(1+C_V)$.

It is interesting at this stage to
compare our results, summarized in
eqs.(\ref{1.17})-(\ref{1.20}), with those presented in \cite{AA}.
Eq.(\ref{1.17}) shows that the $QCD_2$ partition function factorizes
into the partition functions for free fermions, ghosts and
perturbed WZW fields. This factorization (including the remanence
of free fermions) is characteristic of path-integral bosonization
which is always based in the decoupling of the interacting fermions,
which thus become free \cite{RS}-\cite{FNS} (Of course, these free
fermions can in turn
be bosonized in terms of Wess-Zumino fields $-\tilde g$
in ref. \cite{AA}).

In terms of the new variables the BRST symmetry transformation
(\ref{1.11}), (\ref{1.13}) reads
\bear\label{1.21a}
V\delta V^{-1}&=&-\epsilon c_-^{(0)},\nonumber\\
\delta\psi^{(0)}_1&=&\epsilon c_-^{(0)}\psi_1^{(0)},\quad
\delta\psi^{(0)}_2=0\nonumber\\
\delta c_-^{(0)}&=&\frac{\epsilon}{2}\{c_-^{(0)},c_-
^{(0)}\},\quad \delta c_+^{(0)}=0\nonumber\\
\delta b^{(0)}_-&=&\epsilon B_-^{(0)}+\epsilon\Delta_-(V),
\quad \delta b_+^{(0)}=0\ear
where
\bear\label{2.24b}
&&B_-^{(0)}=\psi_1^{(0)}\psi_1^{(0)\dagger}+\{b_-^{(0)},c_-^{(0)}\}
\nonumber\\
&&\Delta_-(V)=-\frac{1}{4e^2}{\D}_-(V)(\partial_+(Vi\partial_-
V^{-1}))-\left(\frac{1+C_V}{4\pi}\right)
Vi\partial_-V^{-1} \nonumber \\
& &
\ear
and
\be\label{1.21b}
{\D}_-(V)=\partial_-+[V\partial_-V^{-1},\quad]\ee
Using (\ref{2.19a}) one readily checks that the
partition function (\ref{1.17}) is invariant under
the above transformation.
The corresponding BRST current,
as obtained via the usual Noether construction, is found to be
(the superscript ``$(B)$'' stands for ``BRST'')
\bear\label{2.28}
&&J_-^{(B)}=tr c^{(0)}_-\left[-\frac{1}{4e^2}D_-(V)\partial_+
\left(Vi\partial_-V^{-1})\right)
-\left(\frac{1+C_V}{4\pi}\right)Vi\partial_-V^{-1}\right.\nonumber\\
&&\left.+\psi_1^{(0)}\psi_1^{(0)\dagger}
+\frac{1}{2}\{b_-^{(0)},c_-^{(0)}\}\right]\ear
with
\be\label{1.22c}
\partial_+J_-^{(B)}=0\ee
Remarkably enough, BRST symmetry leads to a current
which only depends on the variable $x^-$.

It is desirable to rewrite (2.30) in standard form, exhibiting
explicitly its BRST character. To this end we observe
that
\be\label{2.33a}
\Omega_-:=-\frac{1}{4e^2}{\cal D}_-(V)\partial_+(Vi\partial_-V^{-1})-
\left(\frac{1+C_V}{4\pi}\right)Vi\partial_-V^{-1}+j_-\approx 0\ee
with
\bear\label{2.34a}
j_-&=&\psi_1^{(0)}\psi_1^{(0)\dagger}+
\{b_-^{(0)},c_-^{(0)}\}\nonumber\\
&&\partial_+j_-=0\ear
is a constraint of the theory. To see this, we follow the general
ideas outlined in ref. \cite{KS}, ``and gauge'' the partition function
(2.23) with an external field $W_+=i\omega^{-1}\partial_+\omega$,
by making the substitutions
\bear\label{2.35a}
\partial_+&\to&D_+(\omega)=\partial_+-i
W_+\nonumber\\
\partial_+&\to&{\cal D}_+(\omega)=\partial_+-i\left[W_+,\quad \right]
\ear
in the right-hand sector of (2.25) and (2.26),
as well as the substitution
\be\label{2.36a}
\Gamma[V]\to\Gamma[V]-\frac{1}{4\pi}\int tr W_+Vi\partial_-V^{-1}\ee
\be\label{2.37a}
tr\left[\partial_+(V\partial_-V^{-1})\right]^2\to tr
\left[{\cal D}_+(\omega)V\partial_-V^{-1}+i\partial_-W_+\right]^2\ee
in (2.24). Noting that
\be\label{2.38a}
tr\left[{\cal D}_+(\omega)V\partial_-V^{-1}+i\partial_-W_+\right]^2
=tr\left[\partial_+(\omega V)\partial_-(\omega V^{-1})\right]^2\ee
and making use of the Polyakov-Wiegmann identity \cite{PolW}
\be\label{2.39a}
\Gamma[gh]=\Gamma[g]+\Gamma[h]+\frac{1}{4\pi}\int d^2xtr\left(g^{-1}
\partial_+gh\partial_-h^{-1}\right)\ee
one has after a change of integration variable $V\to\omega V$,
\bear\label{2.38a}
&&Z_F^{(0)}\to Z_F^{(0)}e^{-i\Gamma[\omega]}\nonumber\\
&&Z_{gh}^{(0)}\to Z_{gh}^{(0)}e^{-iC_V\Gamma[\omega]}\nonumber\\
&&Z_V\to Z_V e^{i(1+C_V)\Gamma[\omega]}\ear
This shows that the partition function (\ref{1.17}) is unchanged by the

transformations (\ref{2.35a}) to (\ref{2.37a}). From here we derive
the constraint (\ref{2.33a}) by taking the functional derivative of
the gauged partition function with respect to $W_+$, and setting
$W_+=0$. This constraint (Gauss' law)
can be shown to satisfy a Kac-Moody
algebra\footnote{In ref. \cite{Ha} the terminology ``affine Lie
algebra'' is preferred.}
with vanishing central charge; hence, in the
terminology of Dirac \cite{Di}, it is first class.

In terms of the constraint (\ref{2.33a}),
the current (\ref{2.28}) takes

the standard form expected from general
considerations \cite{He}, \cite{FV}:
\be\label{2.39b}
J_-^{(B)}=tr\left[c_-^{(0)}\Omega_--\frac{1}{2}b_-^{(0)}\left\{
c_-^{(0)},c_-^{(0)}\right\}\right]\ee

Since $\Omega_-$ is first class, the corresponding charge
$Q_-^{(B)}$ is nilpotent.

\bigskip
\noindent{\it A second BRST symmetry}

\medskip
As is well known \cite{Ba}, \cite{Ta}, one expects a further BRST
current associated with the change of variables (\ref{1.14}). In fact,
it is easy to see that the partition function (\ref{1.17}) is also
invariant under the transformation
\bear\label{2.35}
&&V^{-1}\delta V=-\epsilon c_+^{(0)}\nonumber\\
&&\delta\psi_1^{(0)}=0,\quad \delta\psi_2^{(0)}=
\epsilon c_+^{(0)}\psi_2^{(0)}\nonumber\\
&&\delta c^{(0)}_-=0,\quad \delta c_+^{(0)}=\frac{\epsilon}{2}
\left\{c_+^{(0)},c_+^{(0)}\right\}\nonumber\\
&&\delta b_-^{(0)}=0,\quad \delta b_+^{(0)}=\epsilon
B_+^{(0)}+\epsilon\Delta_+(V)\ear
with $B_+^{(0)}$ and $\Delta_+(V)$ given by
\bear\label{2.36}
B_+^{(0)}&=&\psi_2^{(0)}\psi_2^{(0)\dagger}+\{b_+^{(0)},
c_+^{(0)}\}\nonumber\\
\Delta_+(V)&=&\frac{1}{e^2}V^{-1}\left(\partial^2_+(Vi\partial_-V^{-1})\right)
V-\frac{1+C_V}{4\pi}V^{-1}i\partial_+V\ear
This transformation law should be compared with the one in (\ref{1.21a}).
The corresponding BRST current obtained via the standard
Noether construction is found to be
\bear\label{2.37}
J_+^{(B)}&=&tr c_+^{(0)}\left[\frac{1}{4e^2}V^{-1}(\partial^2_+(Vi\partial
_-V^{-1})) V-
\frac{1+C_V}{4\pi}V^{-1}i\partial_+V\right.\nonumber\\
&&\left. +\psi_2^{(0)}\psi_2^{(0)\dagger}+\frac{1}{2}
\left\{b_+^{(0)},c_+^{(0)}\right\}\right]\ear
To put this expression into standard form
we observe, following again the method of \cite{KS}, that
gauging with the external field,
\be\label{2.38}
W_-=\frac{i}{e}\omega\partial_-\omega^{-1}\ee
by making the substitutions
\bear\label{2.39}
&&\partial_-\to\partial_--iW_-\nonumber\\
&&\partial_-\to\partial_--i[W_-,\quad ]\ear
in the left-hand sector of (\ref{1.19}) and (\ref{1.20}), we have in
analogy to (\ref{2.38a}),
\bear\label{2.40}
&&Z_F^{(0)}\to Z^{(0)}_Fe^{-i\Gamma[\omega]}\nonumber\\
&&Z_{gh}^{(0)}\to Z^{(0)}_{gh}e^{-iC_V\Gamma[\omega]}\nonumber\\
&&e^{-i(1+C_V)\Gamma[V]}\to e^{-i(1+C_V)(\Gamma[V\omega]-
\Gamma[\omega])}\nonumber\\
&&Vi\partial_-V^{-1}\to V(i\partial_-+W_-)V^{-1}=
(V\omega)i\partial_-(V\omega)^{-1}
\ear
Hence the partition function (\ref{1.17}) is left invariant
by this transformation. From this,  we can derive the constraint
\be\label{2.41}
\Omega_+\equiv\frac{1}{4e}V^{-1}\left[\partial^2_+(Vi\partial_-V^{-1})\right]
V-\frac{(1+C_V)}{4\pi}V^{-1}i\partial_+V+j_+\approx0\ee
where
\bear\label{2.44a}
j_+&=&\psi_2^{(0)}\psi_2^{(0)\dagger}+\left\{b_+^{(0)},
c_+^{(0)}\right\}\nonumber\\
&&\partial_-j_+=0\ear
One readily checks that this constraint is first class (vanishing
central charge).

In terms of $\Omega_+$, the BRST current (\ref{2.37}) is seen to take
the standard form expected from general considerations \cite{He}
\bear\label{2.42}
J_+^{(B)}&=&tr\left[c_+^{(0)}\Omega_+-\frac{1}{2}\left\{
b_+^{(0)},c_+^{(0)}\right\}\right]\nonumber\\
&&\qquad\partial_-J_+^{(B)}=0.\ear
and hence the associated charge is nilpotent.

\section{Non-local decoupled formulation of $QCD_2$}
\setcounter{equation}{0}

The partition function (\ref{1.17}) is
particularly useful in the strong coupling regime. Following
the ideas of ref. \cite{AA}, we now obtain an alternative,
nonlocal representation useful in the weak coupling regime.
To this end we make use of the identity
\bear\label{3.1}
&&\exp
[{\frac{i}{4e^2}\int tr \frac{1}{2}\left[\partial_+(Vi\partial_-V^{-1})
\right]^2}] \nonumber\\
&&=\int{\D}E
\exp[{-i\int
tr\left[\frac{1}{2}E^2+\frac{E}{2e}\partial_+(Vi\partial_-
V^{-1})\right]}]
\ear
and make the change of variable
\be\label{3.2}
\partial_+E =
\left(\frac{1+C_V}{2\pi}\right)\beta^{-1}i\partial_+\beta
\ee
or
\be\label{per}
E = \left(\frac{1+C_V}{2\pi}\right)(\partial_+)^{-1}
\beta^{-1}i\partial_+\beta
\ee
The Jacobian  associated with this change of variables
is
\be\label{3.3}
{\D}E = \exp({-iC_V\Gamma[\beta]}) {\D}\beta
\ee
Making use of the above results, the partition function (\ref{1.17})
reads
\bear
\label{3.5}
Z&=&Z_F^{(0)}Z_{gh}^{(0)} \int{\D}V\int{\D}\beta
\exp\{-i(1+C_V)[\Gamma[V]+\Gamma[\beta]
\nonumber\\
& & -\frac{1}{4\pi}\int tr(\beta^{-1}
\partial_+\beta V\partial_-V^{-1})]\} \nonumber\\
& & \times \exp({ i\Gamma[\beta]})
\exp\left\{{i\left(\frac{1+C_V}{2\pi}\right)^2
e^2\int\frac{1}{2}tr\left[\partial^{-1}_+(\beta^{-1}\partial_+\beta
\right]^2}\right\}
\ear
Now, using the Polyakov-Wiegmann identity (\ref{2.39a})
and making the change of variable $V\to\beta V=\tilde V$,
we are left with
\be\label{3.7a}
Z=Z^{(0)}_FZ^{(0)}_{gh} Z_{\tilde V}Z_\beta\ee
where
\be\label{3.7b}
Z_\beta=\int{\D}\beta \exp\left\{{i\Gamma[\beta]
+i\left(\frac{1+C_V}{2\pi}
\right)^2e^2\int \frac{1}{2}tr\left[\partial_+^{-1}
(\beta^{-1}\partial_+\beta)\right]^2}\right\}
\ee
\be\label{3.8}
Z_{\tilde V}=\int{\D}\tilde V \exp[{-i(1+C_V)\Gamma[\tilde V]}]
\ee
Expression (\ref{3.7a}) agrees with that of ref. \cite{AA}.
\newpage
\noindent
{\it BRST invariance of the $(F-gh-\tilde V)$ sector}

\medskip
The product
$Z_F^{(0)}Z_{gh}^{(0)}Z_{\tilde V}$ is invariant under
the BRST transformations (\ref{1.21a}) and (\ref{2.35}) with
 the substitution $V\to\tilde V$,
and $\Delta_\mp(\tilde V)$ now given by

\be\label{3.9a}
\Delta_-(\tilde V)=-
\left(\frac{1+C_V}{4\pi}\right)
\tilde Vi\partial_-\tilde V^{-1}\ee
and
\be\label{3.9b}
\Delta_+(\tilde V)=-\left(\frac{1+C_V}{4\pi}\right)
\tilde V^{-1}i\partial_+\tilde V\ee
implying the conservation of the corresponding Noether
currents
\be\label{3.10a}
\tilde J_-^{(B)}=tr\ c_-^{(0)}\left[\psi_1^{(0)\dagger}\psi_1^{(0)}
+\frac{1}{2}\left\{b_-^{(0)},c_-^{(0)}\right\}-
\left(\frac{1+C_V}{4\pi}\right)\tilde V\partial_-\tilde V^{-1}
\right]\ee
\be\label{3.10b}
\tilde J_+^{(B)}=tr\ c_+^{(0)}\left[\psi_2^{(0)\dagger}\psi_2^{(0)}
+\frac{1}{2}\left\{b_+^{(0)},c_+^{(0)}\right\}-
\left(\frac{1+C_V}{4\pi}\right)\tilde V^{-1}\partial_+\tilde V
\right]\ee
One easily verifies that the corresponding charges are nilpotent. Indeed,
following again the procedure of ref.\cite{KS}, one
gauges the left- and right-handed
sector as in the conformal sector $(e\to\infty)$
of the local formulation, and shows (with $V$ replaced by $\tilde V$)
that the partition function of the $F-gh-\tilde V$ sector remains
invariant under this gauging.

This implies the existence of the two first-class constraints
\bear\label{13.14c}
&&\tilde\Omega_-\equiv\psi_1^{(0)\dagger}\psi_1^{(0)}+\{b_-^{(0)},c_-^{(0)}\}
-\frac{1+C_V}{4\pi}\tilde Vi\partial_-\tilde V^{-1}\approx0\nonumber\\
&&\tilde\Omega_+\equiv\psi_2^{(0)\dagger}\psi_2^{(0)}+\{b_+^{(0)},c_+^{(0)}\}
-\frac{1+C_V}{4\pi}\tilde V^{-1}i\partial_+\tilde V\approx0\ear
In terms of these constraints, the currents (\ref{3.10a}) and (\ref{3.10b})
take the standard form of BRST currents associated with a first-class
constraint algebra \cite{He}:
\be\label{13.14D}
\tilde J_\pm^{(B)}=tr\left[c_\pm\tilde\Omega^{(0)}_\pm-
\frac{1}{2}b_\pm^{(0)}\left\{c_\pm^{(0)},c_\pm^{(0)}\right\}\right]
\ee
It is important to note that these BRST currents

correspond in the non-local formulation to the currents
(\ref{2.39b}) and (\ref{2.42})
were in the local one.
\bigskip

To conclude this section let us remark that
there exist further BRST-like
symmetries, which, however, are not generated by nilpotent charges.
To take an example, consider the $gh-\tilde V-\beta$ sector.
The partition function $Z_{gh}Z_{\tilde V}Z_\beta$ is readily
seen to be invariant under the symmetry transformation
\newpage
\bear\label{13.15}
&&\tilde V\delta\tilde V^{-1}=-\epsilon c_-^{(0)}\nonumber\\
&&\beta\delta\beta^{-1}=-\epsilon c_-^{(0)}\nonumber\\
&&\delta c_-^{(0)}=\frac{\epsilon}{2}\{c_-^{(0)},c_-^{(0)}\},\quad
\delta c_+^{(0)}=0\nonumber\\
&&\delta b^{(0)}_-=-\epsilon\frac{1+C_V}{4\pi}\tilde V i\partial_-
\tilde V^{-1}+\epsilon\{b_-^{(0)},c_-^{(0)}\}+\epsilon\Delta_-[\beta]\ear
where $\Delta_-[\beta]$ is given by
\be\label{4.5}
\Delta_-(\beta)=\left(\frac{1+C_V}{2\pi}\right)^2e^2
\left[\beta\partial^{-2}_+
(\beta^{-1}i\partial_+\beta)\beta^{-1}\right]+\frac{1}{4\pi}
\beta i\partial_-\beta^{-1}\ee
The corresponding Noether current is found to be
\be\label{3.22}
\tilde J_-=tr\left[ c_-^{(0)}\Omega-\frac{1}{2} b_-^{(0)}\left\{
c_-^{(0)},c_-^{(0)}\right\}\right].\ee
where
\begin{eqnarray}
\label{3.21}
\Omega &\equiv& -\left(\frac{1+C_V}{2\pi}\right)^2e^2\beta
\left(\partial^{-2}_+(\beta^{-1}i\partial_+\beta)\right)\beta^{-1} +
\frac{1}{4\pi}\beta
i\partial_-\beta^{-1}  \nonumber \\
& - & \left(\frac{1+C_V}{4\pi}\right)\tilde V
i\partial_-\tilde
V^{-1}+\left\{ b_-^{(0)},c_-^{(0)}\right\}.
\end{eqnarray}

By gauging the right-handed ghost, and $\tilde V$

sector as in (\ref{2.38a}), (with $V\to\tilde V)$, supplemented
by the transformations
\bear\label{3.21a}
&&\Gamma[\beta]\to\Gamma[\beta]-\frac{i}{4\pi}\int tr W_+
\beta\partial_-\beta^{-1}=\Gamma[\omega\beta]-\Gamma[\beta]\nonumber\\
&&\beta^{-1}\partial_+\beta\to\beta^{-1}(\partial_+-i
W_+)\beta=(\omega\beta)^{-1}\partial_+(\omega\beta)\ear
in the $\beta$-sector, one finds that expression (\ref{3.21})
is constrained to vanish:
\be\label{3.21b}
\Omega\approx 0\ee
As pointed out in \cite{AA}, this constraint is, however,
not first-class with respect to the constraints (\ref{13.14c}), and
as a consequence the BRST charges associated with the currents
(\ref{13.14D}) already represent a complete set.

The role of the constraint (\ref{3.21b}) is best appreciated,
by rewriting the partition function $Z_\beta$ in (\ref{3.7b})
with the aid of an auxiliary field $C_-$ as \cite{AA}
\be\label{3.23}
Z_\beta=\int{\cal D} C_-{\cal D}\beta e^{iS[\beta,C_-]}\ee
where
\be\label{3.24}
S[\beta,C_-]=\Gamma[\beta]+\int tr\left[\frac{1}{2}(\partial_+C_-)^2
+\left(\frac{1+C_V}{2\pi}\right)e(C_-\beta^{-1}i\partial_+\beta)\right]\ee
Gauging $Z_{gh}Z_{\tilde V}Z_\beta$ as described before
$(\partial_+C_-$ and $C_-$ remain unchanged), one arrives at the

constraint:
\be\label{3.25}
\left(\frac{1+C_V}{2\pi}\right)e\beta iC_-\beta^{-1}+
\left(\frac{1+C_V}{4\pi}\right)\tilde V i\partial_-V^{-1}
-\frac{1}{4\pi}\beta i\partial_-\beta^{-1}+\{b_-^{(0)},c_-^{(0)}\}
=0\ee
This constraint determines $C_-$ as a function of the other fields.
Using the equation of motion for $C_-$,
\[\partial_+^2C_-+\left(\frac{2+C_V}{2\pi}\right)e\beta^{-1}\partial_+
\beta=0\]
one then formally arrives at constraint (\ref{3.21b}).

\section{Discussion}
The main objective of this paper was to

further elucidate the interesting analysis of ref. \cite{AA},
by supplementing it with an analysis of the BRST symmetries
as well as the construction of the corresponding BRST currents.

The study of BRST currents is important if one wants
to obtain a complete characterization of the physical
Hilbert space. The formulation of ref. \cite{AA} appears
to be particularly suited for this purpose. Indeed, we have
seen that the conformal invariance of the pure gauged fermionic
partition function, described by a WZW-type theory \cite{Wit},
is broken by the presence of the Yang-Mills action, which implies
the presence of

a coupling constant carrying dimensions.
We have nevertheless seen the BRST currents to be either
left- or right-moving as in a
conformally invariant model. This is
remarkable, and is at the heart of the
claims in ref. \cite{AA}
on the exact integrability of $QCD_2$.
The structure of the physical Hilbert space

as determined by the BRST conditions
is presently under investigation \cite{AR}.

\bigskip
\noindent{\bf Acknowledgement:} One of the authors
(K.D.R.) would like to thank
the Physics Department of the Universidad Nacional de La Plata for the
kind hospitality extended to him, as well as the DAAD for the financial
support making this stay possible. He would also like to thank Elcio
Abdalla for very useful discussions.


\begin{thebibliography}{99}
\bibitem{Schwi} J. Schwinger, Phys. Rev. {\bf 128}, 2425 (1962).
\bibitem{PolW} A. M. Polyakov and P.B. Wiegmann, Phys. Lett. {\bf 131B},
 121 (1983); {\bf 141B},  224 (1984).
\bibitem{Wit} E. Witten, Commun. Math. Phys. {\bf 92}, 455 (1984).
\bibitem{RS} R. Roskies and F. A. Schaposnik {\bf D23},  558 (1981).
\bibitem{GSS} R. E. Gamboa Sarav\'\i , F. A. Schaposnik and J.
E. Solomin, Nucl.
Phys. {\bf B185}, 239 (1982);\\
R. E. Gamboa Sarav\'\i , F. A. Schaposnik and J. E. Solomin,
Phys. Rev. {\bf D30}, 1353 (1984);\\
O. Alvarez, Nucl. Phys. {\bf B216},  125 (1983); ibid. {\bf B238},

61 (1984);\\
R. Roskies, ``Festschrift for Feza Gursey's 60th birthday'',

Symmetries in Particle Physics, eds. I. Bars, A. Chodos, and C.-A.
Tze, Plenum, New York, 1984;\\
K. D. Rothe, Nucl. Phys. {\bf B269}, 269 (1986).
\bibitem{FNS}
E. Fradkin, C. Na\'on and  F. A. Schaposnik,
Phys. Rev. {\bf D36}, 3809  (1987).
\bibitem{AAR} E. Abdalla, M. C. Abdalla and K. Rothe, "Non-Pertubative
Methods in two dimensional Quantum Field Theory", World Sci, 1991.
\bibitem{AA} E. Abdalla and M. C. Abdalla, Int. Jour. Mod. Phys. {\bf A10},
1611 (1995).
\bibitem{Ha} K. Bardakci and M. B. Halpern, Phys. Rev. {\bf D3},
2493 (1971).

\bibitem{Di} P. A. M Dirac, "Lectures on Quantum Mechanics", Yeshiva Univ.
Press., N.Y. 1964; Can. J. Math. {\bf 2}, 129 (1950).
\bibitem{He} M. Henneaux, Phys. Rep. {\bf 126}, 1 (1985).
\bibitem{FV}  E. S. Fradkin and G. A. Vilkovisky, Phys. Lett.
{\bf B55}, 224 (1975); preprint TH 2332 CERN (1977).
\bibitem{Ba} F. Bastianelli, Nucl. Phys. {\bf B361}, 555 (1991).
\bibitem{Ta} Y. Tanii, Mod. Phys. Lett. {\bf A5}, 927 (1990).
\bibitem{KS} D. Karabali and H.J. Schnitzer, Nucl. Phys.
{\bf B329}, 649 (1990).
\bibitem{AR} E. Abdalla and K. D. Rothe, Heidelberg preprint,
in preparation.
\end{thebibliography}
\end{document}